\documentstyle[prl,aps,twocolumn,epsf]{revtex}
\begin{document}
\draft
\twocolumn[
\hsize\textwidth\columnwidth\hsize\csname @twocolumnfalse\endcsname
\title{Non-Appearance of Vortices in Fast Mechanical Expansions
of Liquid $^4$He\\Through the Lambda Transition}

\author{M.E.~Dodd$^1$, P.C.~Hendry$^1$, N.S.~Lawson$^1$,
P.V.E.~McClintock$^1$, C.D.H.~Williams$^2$}

\address{$^1$Department of Physics, Lancaster University,
Lancaster, LA1 4YB, United Kingdom\\
$^2$School of Physics, University of Exeter, Stocker Road, Exeter
EX4 4QL, United Kingdom}

\date{\today}
\maketitle
\widetext

\begin{abstract}

A new experiment has been performed to study the formation of
topological defects (quantized vortices) during rapid quenches of
liquid $^4$He through the superfluid transition, with particular care
taken to minimise vortex creation via conventional hydrodynamic flow
processes. It is found that the generated vortices, if any, are being
produced at densities at least two orders of magnitude less than might
be expected on the basis of the Kibble-Zurek mechanism.

\end{abstract}
]
\narrowtext

\noindent When a physical system passes rapidly enough through a
continuous phase-transition, creation of topological defects is to be
expected \cite{gill} owing to the causal disconnection of separated
regions. This idea was proposed by Kibble \cite{kibble} in connection
with the GUT (grand unified theory) symmetry-breaking phase-transition
of the early universe, and later developed in detail by Zurek
\cite{zureknat,zurekacta,zurekpr} who introduced simple physical
arguments to estimate the density of defects in terms of the rate at
which the system passes through the transition. Zurek also pointed out
that the process of defect production should be generic, applicable in
principle to all continuous phase-transitions, thus opening up the
remarkable possibility of testing aspects of cosmological theories on
the laboratory bench. The first experiments of this kind were performed
on liquid crystals \cite{chuang,bowick} using weakly first-order
phase-transitions. Later, the corresponding experiments were carried
out using the second-order superfluid phase-transitions of liquid
$^4$He \cite{ournature} and liquid $^3$He \cite{bauerle,ruutu}. In all
cases, the generated defect densities were reported as being consistent
with ZurekÕs \cite{zureknat,zurekacta,zurekpr} estimates. In this
Letter we describe an improved version of the $^4$He experiment
\cite{ournature} in which particular care has been taken to minimise
the production of quantized vortices -- the relevant topological
defects in liquid helium -- through ordinary hydrodynamic flow
processes. As we shall see, our new results show no convincing evidence
of any vortex creation at all, but they enable us to place an
approximate upper bound on the initial density of vortices produced via
the Kibble-Zurek mechanism.

The lambda (superfluid) transition in liquid $^4$He provided the basis
of ZurekÕs original proposal \cite{zureknat} for a cosmological
experiment. The underlying idea is very simple. A small isolated volume
of normal (non-superfluid) liquid $^4$He is held under pressure, just
above the temperature $T_{\lambda}$ of the lambda transition. Although
the logarithmic infinity in its heat capacity at $T_{\lambda}$ makes it
impossible to {\it cool} the sample quickly into the superfluid phase,
Zurek pointed out that the pressure dependence of $T_{\lambda}$ meant
that {\it expansion} of the sample would allow it to be taken through
the transition very rapidly. Thus, closely paralleling the processes
through which topological defects (e.g. cosmic strings \cite{vilenkin})
are believed to have been produced in the GUT phase-transition
$10^{-35}$ s after the big bang, huge densities of quantized vortices
\cite{donnelly} should be produced at the transition. The ``bulk
versionÕÕ \cite{zurekacta} of this experiment was subsequently realised
\cite{ournature} in a specially designed expansion apparatus
\cite{ourjltp} in which second-sound attenuation was used to probe the
decaying tangle of vortices created during the expansion. The initial
vortex density, obtained by back-extrapolation to the moment ($t=0$) of
traversing the transition, was $\sim 10^{12}$--$10^{13}$
m$^{-2}$, consistent with the theoretical expectation
\cite{zureknat,zurekacta,zurekpr}.

An unexpected observation in the initial experiments \cite{ournature}
was that small densities of vortices were created even for expansions
that occurred wholly in the superfluid phase, provided that the
starting point was very close to $T_{\lambda}$. The phenomenon was
initially \cite{ourjltp} attributed to vortices produced in thermal
fluctuations within the critical regime, but it was pointed out
\cite{vinenunpub} that effects of this kind are only to be expected
for expansions starting within a few $\mu$K of the transition, i.e.
much closer than the typical experimental value of a few mK. The most
plausible interpretation -- that the vortices in question were of
conventional hydrodynamic origin, arising from nonidealities in the
design of the expansion chamber -- was disturbing, because expansions
starting above $T_{\lambda}$ traverse the same region. Thus some, at
least, of the vortices seen in expansions through the transition were
probably not attributable to the Zurek-Kibble mechanism as had been
assumed. It has been of particular importance, therefore, to undertake
a new experiment with as many as possible of the nonidealities in the
original design eliminated or minimised.

An ideal experiment would be designed so as to avoid all fluid flow
parallel to surfaces during the expansion. This could in principle be
accomplished by e.g. the radial expansion of a spherical volume, or the
axial expansion of a cylinder with stretchy walls. In either of these
cases, the expansion would cause no relative motion of fluid and walls
in the direction parallel to the walls and presumably, therefore, no
hydrodynamic production of vortices. The walls of the actual expansion
chamber \cite{ournature,ourjltp} were made from bronze bellows, thus
approximating the cylinder with stretchy walls. Although there must, of
course, be some flow parallel to surfaces because of the convolutions,
such effects are relatively small. It is believed that the significant
nonidealities, in order of importance, arose from: (a) expansion of
liquid from the filling capillary, which was closed by a needle valve
0.5 m away from the cell; (b) expansion from the shorter capillary
connecting the cell to a Straty-Adams capacitive pressure gauge; (c)
flow past the fixed yoke on which the second-sound transducers were
mounted. In addition (d) there were complications caused by the
expansion system bouncing against the mechanical stop at room
temperature.

\begin{figure}
\epsfxsize=90mm \epsfbox{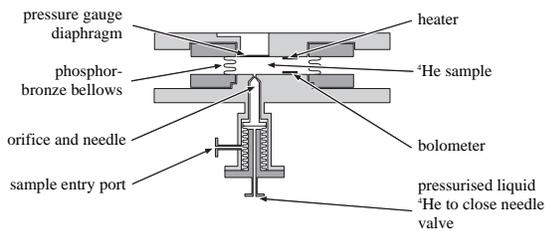}
\caption{Sketch of the improved expansion cell.}
\label{cell} \end{figure}

The new expansion cell, designed to avoid or minimise these problems,
is sketched in Fig.\ \ref{cell}. The main changes from the original
design are as follows: (a) the sample filling capillary is now closed
off at the cell itself, using a hydraulically-operated needle-valve;
(b) the connecting tube to the pressure gauge has been eliminated by
making its flexible diaphragm part of the chamber end-plate; (c) the
second-sound transducers are also mounted flush with the end-plates of
the cell, eliminating any support structure within the liquid; (d) some
damping of the expansion was provided by the addition of a (light motor
vehicle) hydraulic shock-absorber.

\begin{figure}
\epsfxsize=90mm \epsfbox{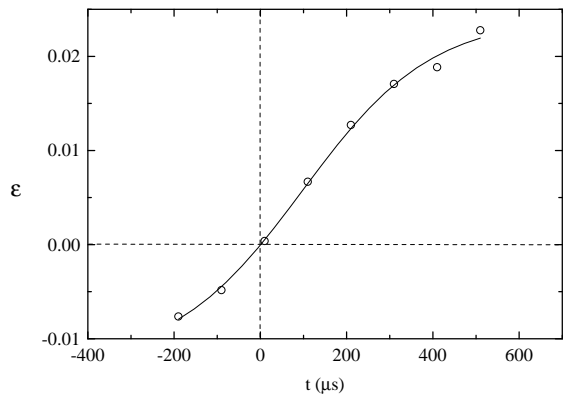}
\caption{Part of a typical quench trajectory, showing
the time evolution of $\epsilon$, the parameter (\ref{epsilon})
specifying distance below the lambda transition. The zero of time $t$
is taken to be the instant when $\epsilon = 0$. The curve is a
guide to the eye, and its slope at $t=0$ enables the quench time
$\tau_Q$ to be determined from (\ref{quench}).} \label{quenchfig}
\end{figure}

The operation of the expansion apparatus and the technique of data
collection/analysis were much as described previously
\cite{ournature,ourjltp}, except that the rate at which the sample passed
through the lambda transition was determined directly by simultaneous
measurements of the position of the pull-rod (giving the volume of the
cell, and hence its pressure) and the temperature of the cell. Distance
from the transition was defined in terms of the parameter
\begin{equation}
\epsilon = \frac{T_{\lambda} - T}{T_{\lambda}}
\label{epsilon}
\end{equation}
Thus the pressure-dependence of $T_{\lambda}$, and the
nonconstant expansion rate, were taken explicitly into account. Part of
a trajectory recorded close to the transition during a typical
expansion is shown in Fig.~\ref{quenchfig}. From such results, the
quench time
\begin{equation}
\tau_Q = \frac{1}{\left(\frac{d \epsilon}{dt}
\right)_{\epsilon = 0}}
\label{quench}
\end{equation}
is easily determined. In the case illustrated, it was $\tau_Q
= 17 \pm 1.0\,\rm ms$. The position of the pull-rod is measured with a ferrite
magnet attached to the rod, moving within a coil fixed to the cryostat
top-plate. There is some evidence of a small time-lag between the
expansion of the cell under the influence of the high-pressure liquid
inside it, and the corresponding movement of the transducer. Thus the
measured $\tau_Q$ may tend to overestimate the quench time slightly.

\begin{figure} \epsfxsize=90mm \epsfbox{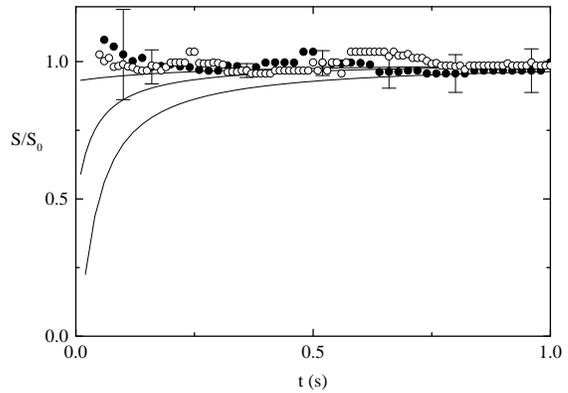}
\caption{Evolution of the second-sound amplitude $S$
with time, following an expansion of the cell at $t=0$ (data points),
normalised by its vortex-free value $S_0$. The open and closed symbols
correspond to signal repetition rates of 100 and 50 Hz respectively,
and in each case the starting and finishing conditions were $\epsilon_i
= -0.032, \epsilon_f = 0.039$. The curves refer to calculated signal
evolutions for different initial line densities, from the bottom, of
$10^{12}, 10^{11}$ and $10^{10}$ m$^{-2}$.} \label{strings}
\end{figure}

Following an expansion though the transition, after $T$, $P$ and the
velocity $c_2$ of second sound have settled at constant values, a
sequence of second-sound pulses is propagated through the liquid. If
the anticipated tangle of vortices is present, the signal may be
expected to grow towards its vortex-free value as the tangle decays and
the attenuation decreases. Signal amplitudes measured just after two
such expansions are shown by the data points of Fig.\ \ref{strings}. It
is immediately evident that, unlike the results obtained from the
original cell \cite{ournature}, there is now no evidence of any
systematic growth of the signals with time or, correspondingly, for the
creation of any vortices at the transition. One possible reason is that
the density of vortices created is smaller than the theoretical
estimates \cite{zureknat,zurekacta,zurekpr}, but we must also consider
the possibility that they are decaying faster than they can be
measured: there is a ``dead periodÕÕ of about 50 ms after the
mechanical shock of the expansion, during which the resultant
vibrations cause the signals to be extremely noisy (which is why the
error bars are large on early signals in Fig.\ \ref{strings}).

The rate at which a tangle of vortices decays in this temperature
range is determined by the Vinen \cite{vinen} equation
\begin{equation}
\frac{dL}{dt} = - \chi_2 \frac{\hbar}{m_4} L^2
\label{vineneq}
\end{equation}
where $L$ is the length of vortex line per unit volume, $m_4$
is the $^4$He atomic mass and $\chi_2$ is a dimensionless parameter.
The relationship between vortex line density and second-sound
attenuation is known \cite{donnelly} from experiments on rotating
helium, and may for present purposes be written in the form
\begin{equation}
L = \frac{6 c_2 {\rm ln} (S_0/S)}{B \kappa d}
\label{atteneq}
\end{equation}
where $S_0$ and $S$ are the signal amplitudes without and
with vortices present, $B$ is a temperature dependent parameter,
$\kappa = h/m_4$ is the quantum of circulation, and $d$ is the
transducer separation.

\begin{figure} \epsfxsize=90mm \epsfbox{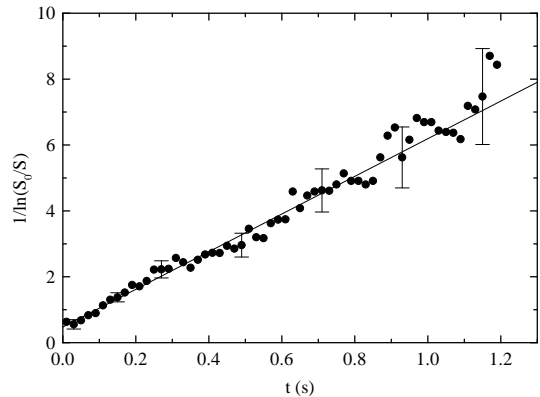}
\caption{Evolution of the second-sound amplitude $S$
following an expansion of the cell with the needle-valve left {\it
open}, thereby causing hydrodynamic vortex creation. The data are
plotted so as to yield a straight line on the basis of (\ref{line}).}
\label{squirt} \end{figure}

Integrating (\ref{vineneq}) and inserting (\ref{atteneq}), one finds
immediately that the recovery of the signal should be of the form
\begin{equation}
\left[{\rm ln} \left( \frac{S_0}{S} \right) \right]^{-1} =
\frac{6c_2}{\kappa B d} \left(\chi_2 \frac{\kappa}{2\pi}t + L_i^{-1} \right)
\label{line}
\end{equation}
Of the constants in (\ref{line}), all are known except
$\chi_2$ and $B$ which seem not to have been measured accurately within
the temperature range of interest. We therefore performed a subsidiary
experiment, deliberately creating vortices by conventional means and
then following their decay by measurements of the recovery of the
second-sound signal amplitude. This was accomplished by leaving the
needle-valve open, so that $\sim 0.2$ cm$^3$ of liquid from the dead
volume outside the needle-valve actuator-bellows was forcibly squirted
into the cell during an expansion. As expected, large densities of
vortices were created. Plots of $[{\rm ln}(S/S_0)]^{-1}$ as a function
of $t$ yielded straight lines within experimental error in accordance
with (\ref{line}), as shown in Fig.\ \ref{squirt}. By measurement of
the gradient, and comparison with (\ref{line}), for a number of such
plots it was possible to determine $\chi_2/B$ as a function of $T$ and
$P$. Fuller details willl be given elsewhere, but it was found that
$\chi_2/B$ did not exhibit a strong temperature dependence and could be
approximated by $\chi_2/B = 0.004 \pm 0.001$ within the range of
interest where $0.02 < \epsilon < 0.06$.

The measured value of $\chi_2/B$ was then used to calculate the evolution
of $S/S_0$ with time for different values of $L_i$, yielding the curves
shown in Fig.\ \ref{strings}.  From the $\tau_Q$ measured
from the gradient in Fig.\ \ref{quenchfig}, and ZurekÕs estimate of
\begin{equation}
L_i = \frac{1.2 \times 10^{12}}{(\tau_Q/100 \, {\rm ms})^{2/3}}
\quad \quad [{\rm m}^{-2}]
\label{density}
\end{equation}
we are led to expect that $L_i \approx 4 \times 10^{12}$
m$^{-2}$. A comparison of the calculated curves and measured data in
Fig.\ \ref{strings} shows that this is plainly not the case. In fact,
the data suggest that  $L_i$ is smaller than the expected value by at
least two orders of magnitude.

There are a number of comments to make on this null result which, in
the light of the apparently positive outcome from the earlier
experiment \cite{ournature}, came as a considerable surprise. First,
ZurekÕs estimates of $L_i$ were never expected to be accurate to better
than one, or perhaps two, orders of magnitude, and his more recent
estimate \cite{zureklater} suggests somewhat lower defect densities. So
it remains possible that his picture
\cite{zureknat,zurekacta,zurekpr,zureklater} is correct in all
essential details, and that the improved experiment with faster
expansions now being planned will reveal evidence of the Kibble-Zurek
mechanism in action in liquid $^4$He. Secondly, it must be borne in
mind that (\ref{vineneq}), and the value of $\chi/B$ measured from the
data of Fig.\ \ref{squirt}, refer to hydrodynamically generated vortex
lines. Vorticity generated in the nonequilibrium phase transition may
be significantly different, e.g.\ in respect of its loop-size
distribution \cite{gaw}, may therefore decay faster, and might
consequently be unobservable in the present experiments. Thirdly
however, it seems surprising that the $^3$He experiments
\cite{bauerle,ruutu} appear to give good agreement with ZurekÕs
original estimates \cite{zureknat,zurekacta,zurekpr} whereas the
present experiment shows that they overestimate $L_i$ by at least two
orders of magnitude. It is not yet known for sure why this should be,
but an interesting explanation of the apparent discrepancy is being
proposed \cite{karra} by Karra and Rivers.

We acknowledge valuable discussions or correspondence with S.N.\
Fisher, A.J.\ Gill, R.A.M.\ Lee, R.J.\ Rivers, W.F.\ Vinen, G.A.\
Williams and W.H.\ Zurek. The work was supported by the Engineering and
Physical Sciences Research Council (U.K.), the European Commission and
the European Science Foundation.

\filbreak

\end{document}